\documentstyle[aps]{revtex}

\begin{document}

\title{Linear and nonlinear optical characteristics
of the Falicov-Kimball model}

\author{T. Portengen, Th. \"{O}streich, and L. J. Sham}
\address{Department of Physics, University of California
San Diego, La Jolla, California 92093-0319}

\date{31 March 1996}

\maketitle

\begin{abstract}
We calculate the linear and nonlinear optical properties
of the Falicov-Kimball model for a mixed-valent system within
the self-consistent mean-field approximation.
Second-harmonic generation can only occur if
the mixed-valent state has a built-in coherence
between the itinerant $d$\/-electrons and the localized
$f$\/-holes. By contrast, second-harmonic generation cannot
occur for solutions of the model with $f$\/-site occupation
as a good quantum number. As an experimental test of coherence
in mixed-valent compounds we propose a measurement of the
dynamic second-order susceptibility.
\end{abstract}


\narrowtext

The Falicov-Kimball~\cite{Falicov} (FK) model has been
used extensively for the mixed-valent compounds, heavy
fermion systems, and associated metal-insulator transitions.
The FK model accounts for a band of itinerant $d$\/-electrons
and localized $f$\/-orbitals and intrasite Coulomb interaction
between the $d$\/- and $f$\/-electrons. A $d$\/-$f$ hybridization
term may or may not be added to the model. The theoretical
solutions for the ground state of the FK model can be divided
into two classes. On the one hand, solutions that treat the
occupation of an $f$\/-electron on a site as a good quantum
number~\cite{Freericks,Farkasovsky} do not have a built-in
coherence between $d$\/-electrons and $f$\/-holes.
On the other hand, solutions such as the self-consistent
mean-field solution~\cite{Leder} and the electronic
polaron~\cite{Liu} do have a built-in coherence between
$d$\/-electrons and $f$\/-holes.

We report here the nonlinear optical responses of these two
classes of solutions. Solutions of the model with a built-in
coherence can sustain second-harmonic generation.
Solutions with classical $f$\/-electron site distributions cannot.
Therefore, we propose the measurement of the second-order
susceptibility of a mixed-valent compound as a test to distinguish
between these theories. The existence of such second-harmonic
generation due to coherence in the ground state would,
of course, be of interest in its own right as
a manifestation of strong electron correlation.

Four-Wave-Mixing (FWM) spectroscopy has become a powerful
tool for studying coherence in semiconductor systems
\cite{Kim}.
In a three-beam FWM experiment, two incoming beams of wavevectors
${\bf k}_{1}$ and ${\bf k}_{2}$ set up a transient polarization
grating. The third incoming beam of wavevector ${\bf k}_{3}$
diffracts off the grating to produce the outgoing signal in
the direction ${\bf k}_{4}={\bf k}_{3}+{\bf k}_{2}-{\bf k}_{1}$.
Being a third-order process, FWM is allowed in media with or
without inversion symmetry. We pose the question: what
happens if the state being probed already has a polarization
built into it instead of being created artificially
 by optical pumping?
An example of such a system is the self-consistent
mean-field (SCMF) solution of the FK model resulting in the
Bose-Einstein condensation of $d$\/-$f$ excitons.

As shown below, the built-in polarization leads to a nonlinear
optical response to {\em second} order in the external field.
The built-in polarization replaces one of the external fields
of the three-beam FWM experiment. The mixed-valent system has
a nonvanishing susceptibility $\chi^{(2)}(2\omega,\omega,\omega)$
for second-harmonic generation. In crystals with inversion
symmetry, second-harmonic generation is forbidden under the
electric-dipole approximation. In the mixed-valent system,
the built-in polarization breaks the inversion symmetry,
allowing second-harmonic generation to take place. The built-in polarization
also means that the system is ferroelectric.  Our calculation of the
concomitant
dielectric behavior will be reported elsewhere.  There are reports of usually
large dielectric constants in mixed valent semiconductors \cite{goto}.  Because
of the problem of residual carriers in dielectric measurements, we feel that
the nonlinear optical effect might be a clearer test.

We present a calculation of the linear and second-harmonic
susceptibilities of a model mixed-valent system within the SCMF
approximation. The magnitude of the second-harmonic
output signal is directly proportional to the built-in
coherence $\Delta$. The Coulomb interaction between the optically
excited quasiparticles greatly enhances the second-harmonic
conversion efficiency at $\omega=\Delta$ (one half the energy
gap $2\Delta$).

Ignoring the electron spin, the FK Hamiltonian is
\begin{eqnarray}
\label{eq:FKH}
H & = &
\sum_{\bf k} \varepsilon_{\bf k} d^{\dagger}_{\bf k} d_{\bf k} +
E'_{f} \sum_{\bf k} f^{\dagger}_{\bf k} f_{\bf k} +
\sum_{\bf k} (V_{\bf k} d^{\dagger}_{\bf k}f_{\bf k} + \mbox{ h.c.})
\nonumber \\ & & +
\frac{U}{N} \sum_{\bf k,k',q} d^{\dagger}_{\bf k +q} d_{\bf k }
                              f^{\dagger}_{\bf k'-q} f_{\bf k'} .
\end{eqnarray}
Here $d^{\dagger}_{\bf k}$ ($f^{\dagger}_{\bf k}$) creates a
$d$\/-($f$\/-)electron of momentum ${\bf k}$ and energy
$\varepsilon_{\bf k}$ ($E'_{f}$). The parameters $U$ and $V_{\bf k}$
are the direct interaction and the hybridization between the
$d$\/- and $f$\/-electrons, and $N$ is the number of sites.
We consider a model system with a $d$\/-band and $f$\/-level
arising from $d$\/- and $f$\/-orbitals on the same site. The
$d$\/-band has bandwidth $2W$ and a constant density of states
$\rho_{0}=1/(2W)$.

The SCMF solution is analogous to the BCS theory of superconductivity
except that the pairing now occurs between a $d$\/-electron of
momentum ${\bf k}$ and a $f$\/-{\em hole} of momentum $-{\bf k}$.
The ground state is
\begin{math}
|\psi\rangle = \prod_{\bf k} (u_{\bf k} + v_{\bf k}
                       d^{\dagger}_{\bf k}f_{\bf k})|0\rangle ,
\end{math}
where $|0\rangle$ is the state with no {\em f}\/-holes (the normal
state), and $v_{\bf k}$ ($u_{\bf k}$) is the probability amplitude
for the pair state $({\bf k},-{\bf k})$ to be occupied (unoccupied).

A key feature of $|\psi\rangle$ is that it is a state of broken
inversion symmetry. If the crystal is invariant under inversion
with respect to a $d$\/-$f$ site, the inversion symmetry
is broken by the pairing of $d$\/-states of even parity with
$f$\/-states of odd parity. Applying the inversion $\hat{J}$ on
$|\psi\rangle$ yields the state $\hat{J}|\psi\rangle = \prod_{\bf k}
(u_{\bf k}^* - v_{\bf k}^* d^{\dagger}_{\bf k}f_{\bf k})|0\rangle$, which is
orthogonal to $|\psi\rangle$ and has the same energy except in the case $U=0$
when the two states are the same. The degenerate states $|\psi\rangle$ and
$\hat{J}|\psi\rangle$ have built-in polarizations of opposite directions, for
the polarization operator $\hat{\bf P}=(
\sum_{\bf k}
\mbox{\boldmath $\mu$}
 d^{\dagger}_{\bf k}f_{\bf k}+{\rm h.c.} )/\Omega$
where $\Omega$ is the system volume.
We take the interband dipole matrix element
 \mbox{\boldmath $\mu$} to be independent of ${\bf k}$.
The correct ground state is selected by lifting the degeneracy
with an infinitesimal external electric field ${\bf E}$, and
choosing the lower energy state. The consequent breaking of the
inversion symmetry is what allows second-harmonic generation
to take place.

The built-in polarization defines a direction
in space, which we call the $z$\/-axis. (Without crystal-field
terms, the $z$\/-axis has no definite orientation with respect
to the crystal axes.) Since $\mu_{z}$ is real,
   $P^{(0)}_{z} = N \mu_{z} (\Delta+\Delta^{*})/(\Omega U)$,
where $\Delta$ is the built-in coherence.
The built-in coherence is determined self-consistently from
\begin{equation}
\label{eq:GAP}
\Delta = \frac{U}{N}
\sum_{\bf k} \frac{\Delta-V_{\bf k}}{2 E_{\bf k}} ,
\end{equation}
where
 $2E_{\bf k}=\sqrt{(\varepsilon_{\bf k}-E_{f})^{2}+
4|\Delta-V_{\bf k}|^{2}}$
is the quasi-electron-hole pair excitation energy~\cite{Note}.
Eq.~(\ref{eq:GAP}) is Eq.~(11) of Ref.~\onlinecite{Leder}
with a ${\bf k}$\/-dependent hybridization.
If the crystal is invariant under inversion,
the hybridization must satisfy $V_{-\bf k}=-V_{\bf k}$.
Then, since $V_{\bf k}$ is purely imaginary and odd in ${\bf k}$,
the imaginary part of $\Delta$ vanishes due to the cancellation
of terms with $\pm {\bf k}$. The real part of $\Delta$ is given by
the BCS gap equation
\begin{equation}
\Delta = \frac{U}{N} \sum_{\bf k} \frac{\Delta}{2 E_{\bf k}} ,
\end{equation}
with $2E_{\bf k} = \sqrt{( \varepsilon_{\bf k}-E_{f})^{2} +
                          4\Delta^{2}+4|V_{\bf k}|^{2}}$.
Calculation shows that sufficiently strong $V_{\bf k}$ can destroy the gap.
In the following we consider the limit where $V_{\bf k}$ is neligible.

$\Delta$ is the order parameter of the valence transition.
When the $f$\/-level is far below the $d$\/-band, the
system is in the normal state ($\Delta=0$). As the $f$\/-level is
moved upward past a critical value (in a real material this is done
by applying pressure or alloying), the system undergoes a transition
into the mixed-valent state ($\Delta>0$). In the mixed-valent state,
the $f$\/-level occupancy $n_{f}$ lies between zero and one.
$\Delta$ reaches a maximum at the half-filling point $E_{f}=0$.
Electron-hole symmetry requires $\Delta(-E_{f}) = \Delta(E_{f})$
and $n_{f}(-E_{f})=1-n_{f}(E_{f})$.

We first consider the linear absorption spectrum of
the mixed-valent system. The SCMF solution predicts an
energy gap~$2\Delta$. Far-infrared optical
measurements~\cite{Wachter,Molnar,Batlogg}, as well as
electron tunneling experiments~\cite{Guntherodt}, show
an energy gap of several meV in a number of
mixed-valent compounds. The crucial difference between the
superconductor and the mixed-valent system is this: in the
superconductor, the pairing occurs between two {\em electrons},
whereas in the mixed-valent system, the pairing occurs between
an {\em electron} and a {\em hole}. This has important consequences
for the coherence factors that enter the response of both systems
to different external probes. For example, the coherence factor
entering the {\em optical} absorption of the mixed-valent system
is the same as the coherence factor entering the {\em acoustic}
attenuation of the superconductor.
The interaction of the mixed-valent system with the electromagnetic
field is treated in the electric-dipole approximation. Only the
component of the electric field along the symmetry-breaking
$z$\/-axis couples to the channel in which the pairing takes place.
We ignore the response of the remaining optical channels.
Second-harmonic generation can only occur
in the symmetry-breaking channel.

We have calculated the linear susceptibility $\chi^{(1)}_{zz}$ both
from the Kubo formula and from the optical Bloch equations. The
pseudo-spin picture gives a nice physical description of the linear
and nonlinear responses as precessional modes of the pseudo-spin
${\bf S}_{\bf k}$. For a given ${\bf k}$, the pseudo-spin corresponds
to the $d_{\bf k}$\/- and the $f_{\bf k}$\/-state as a two-level
system. The equations of motion for the pseudo-spin are
the optical Bloch equations
\begin{equation}
\label{eq:BLOCH}
\dot{\bf S}_{\bf k} = ({\bf H}_{\bf k}-{\bf M}_{\bf k}) \times
                       {\bf S}_{\bf k} .
\end{equation}
Here ${\bf H}_{\bf k} = (-2\mu_{z}E_{z},0,\varepsilon_{\bf k}-E'_{f})$
is the external ``magnetic'' field, and ${\bf M}_{\bf k} = \frac{U}{N}
\sum_{\bf k} {\bf S}_{\bf k}$ is the pseu\-do-mag\-ne\-ti\-za\-tion.
The symbol $\times$ means the vector cross product.

To calculate the linear susceptibility we expand ${\bf S}_{\bf k}$,
${\bf H}_{\bf k}$ and ${\bf M}_{\bf k}$ to first order in the
electric field $E_{z}$. From Eq.~(\ref{eq:BLOCH}) to zeroth order
we obtain an equation for the stationary pseudo-spin
${\bf S}^{(0)}_{\bf k}$. The built-in coherence tilts
${\bf S}^{(0)}_{\bf k}$ away from the negative $z$\/-axis.
The tilting angle is $\theta_{\bf k} =
\arccos(v^{2}_{\bf k}-u^{2}_{\bf k})$. For $\Delta$ real,
${\bf S}^{(0)}_{\bf k}$ lies in the $x$\/-$z$ plane.
The electric field causes the pseudo-spin to precess around
the stationary direction. With the precession axis tilted away
from the $z$\/-axis, the field causes variations in all three
Cartesian components of ${\bf S}_{\bf k}$. A simpler description
is obtained in the spherical polar coordinate system.
The stationary direction is the radial unit vector
${\bf e}_{r}$. The precession is decomposed into components along
the polar and azimuthal unit vectors ${\bf e}_{\theta}$ and
${\bf e}_{\phi}$. The equations of motion for the polar and
azimuthal components $S^{(1)}_{\theta,\bf k}$
and $S^{(1)}_{\phi,\bf k}$ are
\begin{eqnarray}
\label{eq:MODE_1}
\dot{S}^{(1)}_{\theta,\bf k} - 2E_{\bf k} S^{(1)}_{\phi,\bf k}
                                        + M^{(1)}_{\phi,\bf k}
& = & F^{(1)}_{\theta,\bf k} , \\
\label{eq:MODE_2}
\dot{S}^{(1)}_{\phi,\bf k} + 2E_{\bf k} S^{(1)}_{\theta,\bf k}
                                      - M^{(1)}_{\theta,\bf k}
& = & F^{(1)}_{\phi,\bf k} .
\end{eqnarray}
The driving terms are $F^{(1)}_{\theta,\bf k} = 0$ and
$F^{(1)}_{\phi,\bf k} = 2 \mu_{z}E_{z} \cos \theta_{\bf k}$.
The linear susceptibility is $\chi^{(1)}_{zz} = P^{(1)}_{z}/E_{z}$,
where $P^{(1)}_{z} = ( N \mu_{z}
 \sum_{\bf k} S^{(1)}_{\theta,\bf k} \cos\theta_{\bf k} )/\Omega$
 is the linear polarization.

For a separable interaction potential, Eqs.~(\ref{eq:MODE_1})
and (\ref{eq:MODE_2}) can be solved analytically.
The linear susceptibility is
\begin{equation}
\label{eq:CHIONE}
\chi^{(1)}_{zz} = \frac{2 N \mu_{z}^{2}}{\Omega U}
\left( \frac{A(\omega)}{ (\omega^{2}-4\Delta^{2})
         A^{2}(\omega)-B^{2}(\omega)} - 1
\right),
\end{equation}
where
\begin{eqnarray}
A(\omega) & = & \frac{U}{N} \sum_{\bf k}
\frac{1}{2 E_{\bf k}(\omega-2 E_{\bf k})
                    (\omega+2 E_{\bf k})} , \\
B(\omega) & = & \frac{U}{N} \sum_{\bf k}
\frac{\varepsilon_{\bf k}-E_{f}}
       {2 E_{\bf k}(\omega-2 E_{\bf k})
                   (\omega+2 E_{\bf k})} .
\end{eqnarray}
For the simple model system, $A(\omega)$ and $B(\omega)$
can be expressed in terms of elementary functions.
The poles of $\chi^{(1)}_{zz}$ give the collective excitation
energies of the mixed-valent system. The denominator in
Eq.~(\ref{eq:CHIONE}) vanishes when
$(\omega^{2}-4\Delta^{2})A^{2}(\omega)-B^{2}(\omega)=0$.
The zero-frequency Goldstone mode is a consequence of the
arbitrariness of the phase of $\Delta$ (in the absence of
hybridization). In the pseudo-spin picture, the Goldstone
mode corresponds to rotating all pseudo-spins
around the $z$\/-axis over the same angle $\phi$.
 Since this does not change the total energy, the Goldstone
 mode has zero frequency. In Eq.~(\ref{eq:CHIONE}),
 the Goldstone mode does not appear to
contribute to the linear optical response
since the pole at $\omega=0$ is canceled by a factor of $\omega$ in the
numerator.  There are no exciton-like
collective modes within the energy gap. When $\omega<2\Delta$,
the functions $A(\omega)$ and $B(\omega)$ are purely real, so
$(\omega^{2}-4\Delta^{2})A^{2}(\omega)-B^{2}(\omega)<0$.

The absorption spectrum is given by the imaginary part of
$\chi^{(1)}_{zz}$.
When the $f$\/-level lies inside the $d$\/-band ($|E_{f}|\leq W$),
the absorption spectrum has a threshold singularity at
$\omega=2\Delta$. When $|E_{f}|<W$, the
singularity is $\epsilon^{-1/2}\theta(\epsilon)$, and
when $|E_{f}|=W$ the singularity is
$\epsilon^{-1/2}\ln^{-2}(\epsilon)\theta(\epsilon)$,
where $\epsilon=\omega-2\Delta$. When the $f$\/-level lies
outside the $d$\/-band the singularity is cut off because
the energy gap is larger than $2\Delta$. The singularity is
due to the final-state Coulomb interaction between the optically
excited quasiparticles. In the single-quasiparticle picture,
the absorption spectrum rises continuously from zero according
to $\epsilon^{1/2}\theta(\epsilon)$.
The singularity is {\em not} an artifact of the simple model,
and should be observable in real materials.

We calculate the second-harmonic susceptibility $\chi^{(2)}_{zzz}$
from the optical Bloch equations by expanding the pseudo-spin and
the pseudo-magnetization to {\em second} order in the perturbing
electric field $E_{z}$. The equations of motion for the second-order
components $S^{(2)}_{\theta,\bf k}$ and $S^{(2)}_{\phi,\bf k}$ have
the same form as Eqs.~(\ref{eq:MODE_1}) and (\ref{eq:MODE_2}), except
with more complicated driving terms.
A very important observation is that all driving terms are
directly proportional to $\Delta$. When $\Delta=0$,
the second-harmonic susceptibility vanishes identically.

In addition, the se\-cond-or\-der fluc\-tu\-a\-tions have a
non\-zero ra\-di\-al component $S^{(2)}_{r,\bf k}$. The motion
is no longer a regular precession: the pseudo-spin {\em nutates}
during the precession. (Nutation is the up-and-down motion of
the precession axis.) The nutation frequency is twice the
precession frequency.
The second-harmonic susceptibility is $\chi^{(2)}_{zzz} =
P^{(2)}_{z}/E^{2}_{z}$, where $P^{(2)}_{z} =
 [N \mu_{z}
  \sum_{\bf k}(S^{(2)}_{\theta,\bf k} \cos \theta_{\bf k}
+ S^{(2)}_{r,\bf k} \sin \theta_{\bf k})]/\Omega$
is the second-order polarization.

For a separable interaction potential, an analytic solution for the
second-harmonic response is possible in principle. However,
the large number of driving terms presents a considerable
challenge. We have instead approached the problem numerically.
This is done in analogy with the classical
mechanics treatment of forced oscillations. One first solves
for the motion in normal coordinates, and then takes linear
combinations to obtain the motion in the original coordinates.
The results of the calculation are shown in Fig.~\ref{fig:CHITWO}.
The figure shows the amplitude
$|\chi^{(2)}_{zzz}(2\omega,\omega,\omega)|$ of the second-harmonic
susceptibility as a function of the photon energy $\omega$, for
various values of $E_{f}$. The important features are:
(1) The second-harmonic amplitude is directly
proportional to the amount of coherence $\Delta$ built into the
mixed-valent system. (2) When the $f$\/-level lies inside the
$d$\/-band, the second-harmonic conversion efficiency is strongly
enhanced at $\omega=\Delta$, and less strongly at
$\omega=2\Delta$. The first feature shows that second-harmonic
generation can be used as a test of the validity of the SCMF
solution in real mixed-valent materials. The second feature
distinguishes the single-quasiparticle treatment of the
second-harmonic response
from the self-consistent mean-field treatment. Like the threshold
singularity in the case of linear reponse, the enhancement of the
second-harmonic conversion efficiency is due to the final-state
Coulomb interaction between the optically excited quasiparticles.

As an experimental test of coherence in mixed-valent compounds
we propose a measurement of their second-harmonic susceptibility
$\chi^{(2)}(2\omega,\omega,\omega)$.
Consider for example ${\rm SmB}_{6}$.
The crystal structure of ${\rm SmB}_{6}$ has cubic symmetry,
with ${\rm B}_{6}$ octahedra at the body center, and ${\rm Sm}$
ions at the corners of a conventional bcc unit cell with lattice
constant $a=4.13$\,\AA. The crystal has inversion symmetry at
the bcc lattice points. Through measurements of the ionic radius,
the valence of the ${\rm Sm}$ ion in ${\rm SmB}_{6}$ is found to
be 2.53, almost halfway between 2 and 3, so that the $f$\/-level
lies near the center of the conduction band.
The measured far-infrared absorption
 spectrum~\cite{Wachter,Molnar,Batlogg} of ${\rm SmB}_{6}$
 can be interpreted in accordance with the SCMF solution.
In Figure~\ref{fig:CHIONE} we compare the mean-field and
single-quasiparticle results for the linear susceptibility to
experimental data on ${\rm SmB}_{6}$
taken from Ref.~\onlinecite{Wachter}.
The data show an energy gap
around $2\Delta=4\,{\rm meV}$, and a sharp peak at threshold.
The mean-field theory fits the data very well in the threshold
region, whereas the single-quasiparticle theory gives a
qualitatively wrong threshold behaviour. Away from threshold,
discrepancies between mean-field theory and experiment occur
 because of our simple model density of states.
Further experimental indication for the validity of the SCMF
 solution
in ${\rm SmB}_{6}$ is provided by the electron tunneling
spectrum~\cite{Guntherodt}, which can be explained by analogy with
Giaever tunneling in superconductors.

The second-harmonic generation for the other types of solutions
of the FK model is now discussed briefly. If the $f$\/-occupancy at
each site is a good quantum number ($0$ or $1$), any solution
\cite{Freericks,Farkasovsky}, homogeneous or inhomogeneous, will not
give rise to second-harmonic generation. For the electronic polaron
solution~\cite{Liu}, the nonzero coherence yields a second-harmonic
generation. We found $\Delta=3.28 W$ for the parameter values given
in Ref.~\onlinecite{Liu}. To distinguish between the exciton
condensation solution and the electronic polaron requires an
investigation of the quantitative difference of their linear and
nonlinear optical spectra. Such a theoretical study will be left
for the future.

In conclusion, we calculated the linear and nonlinear optical
characteristics of the Falicov-Kimball model within the SCMF
approximation. We found that the second-harmonic susceptibility is
directly proportional to the amount of coherence $\Delta$ built into
the mixed-valent system. We also found that the final-state Coulomb
interaction leads to a threshold singularity in the absorption spectrum,
and strongly enhances the second-harmonic conversion efficiency at
$\omega=\Delta$. As an experimental test of the validity of the SCMF
solution in real mixed-valent materials we propose a measurement of
$\chi^{(2)}(2\omega,\omega,\omega)$.

LJS wishes to thank Dr. M. B. Maple and Dr. S. H. Liu
for stimulating conversations. This work was supported
in part by NSF Grant No. DMR 94-21966 and in part by
the Deutsche Forschungsgemeinschaft (DFG).

\begin{figure}
\caption{\label{fig:CHITWO}
 Amplitude $|\chi^{(2)}_{zzz}(2\omega,\omega,\omega)|$
 of the se\-cond-har\-mo\-nic susceptibility as a function of the
photon energy $\omega$, for various values of the $f$\/-level
$E_{f}$. The dash-dotted line shows the phase of
 $\chi^{(2)}_{zzz}(2\omega,\omega,\omega)$ for $E_{f}=-1.0\,W$.
  The Coulomb repulsion is
 $U=3.0\,W$ and the hybridization is $V_{\bf k}=0$.
 The amplitude is given in units of $N \mu_{z}^{3}/(2\Omega W^{2})$.
 For the parameter values given for the solid line in Fig.~2,
 $N \mu_{z}^{3}/(2\Omega W^{2})=82\,{\rm nm\,V^{-1}}$.}
\end{figure}

\begin{figure}
\caption{\label{fig:CHIONE}
Comparison of the mean-field (solid line) and
 single-quasiparticle (dash-dotted line) results for
 the imaginary part of the linear susceptibility
 $\chi^{(1)}_{zz}(\omega)$ of SmB$_{6}$
 to experimental data taken from Ref.~10 (diamonds).
 The $f$\/-level is $E_{f}=0$, the bandwidth is $W=40\,{\rm meV}$,
 the Coulomb repulsion is $U=0.38\,W$, and the hybridization is
   $V_{\bf k}=0$.
 The photon energy $\omega$ was given a small imaginary part
  $\delta = 0.01\,W$.
 The interband dipole matrix element is $\mu_{z}=4.4\,ea_{0}$
  for the solid line, and  $\mu_{z}=5.0\,ea_{0}$ for the
 dash-dotted line.}
\end{figure}


\begin{thebibliography}{99}
\bibitem{Note}
We shall use the renormalized $f$\/-level $E_{f}=E'_{f}+U(1-2n_{f})$
as the materials parameter.
\end{thebibliography}
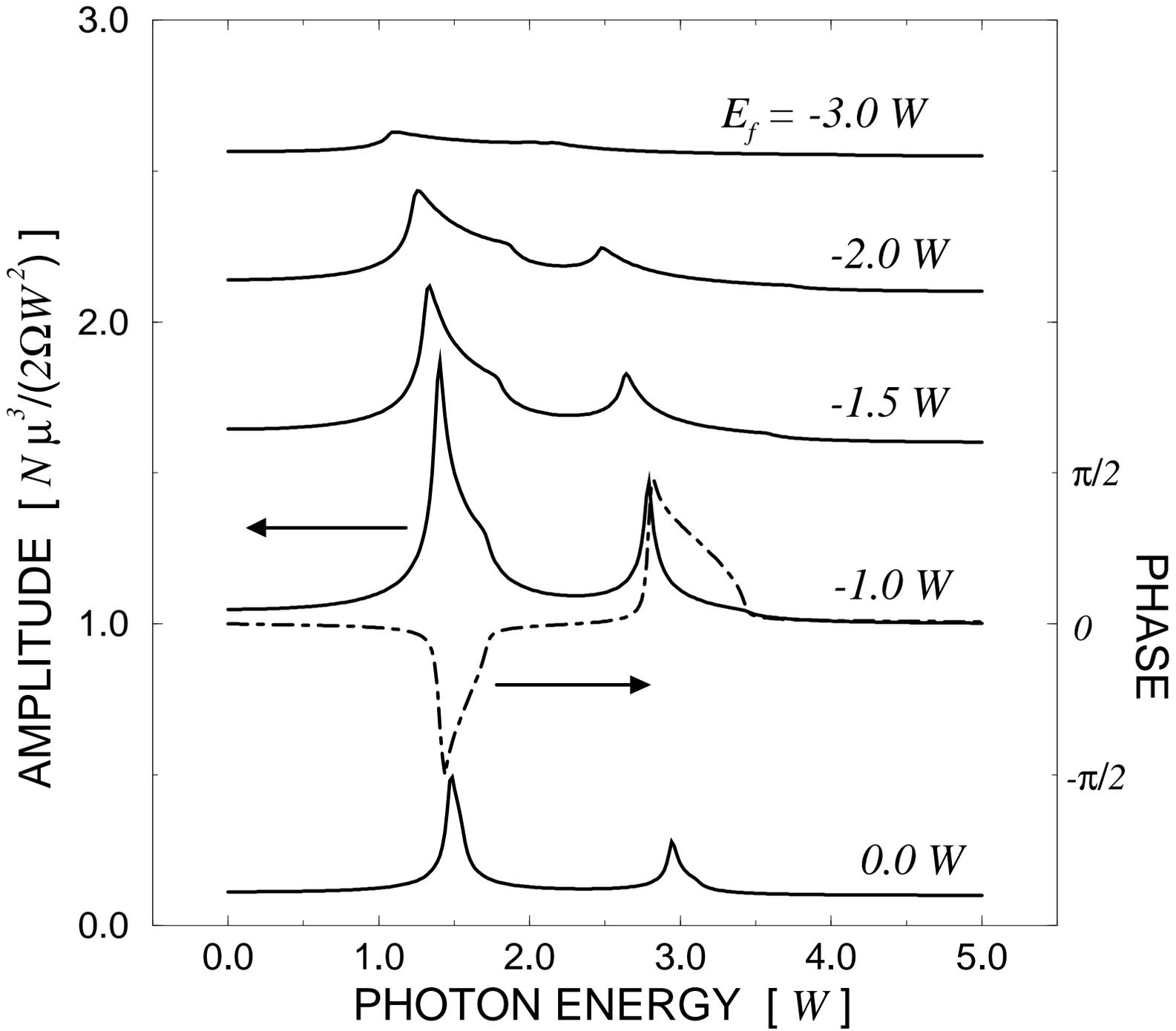
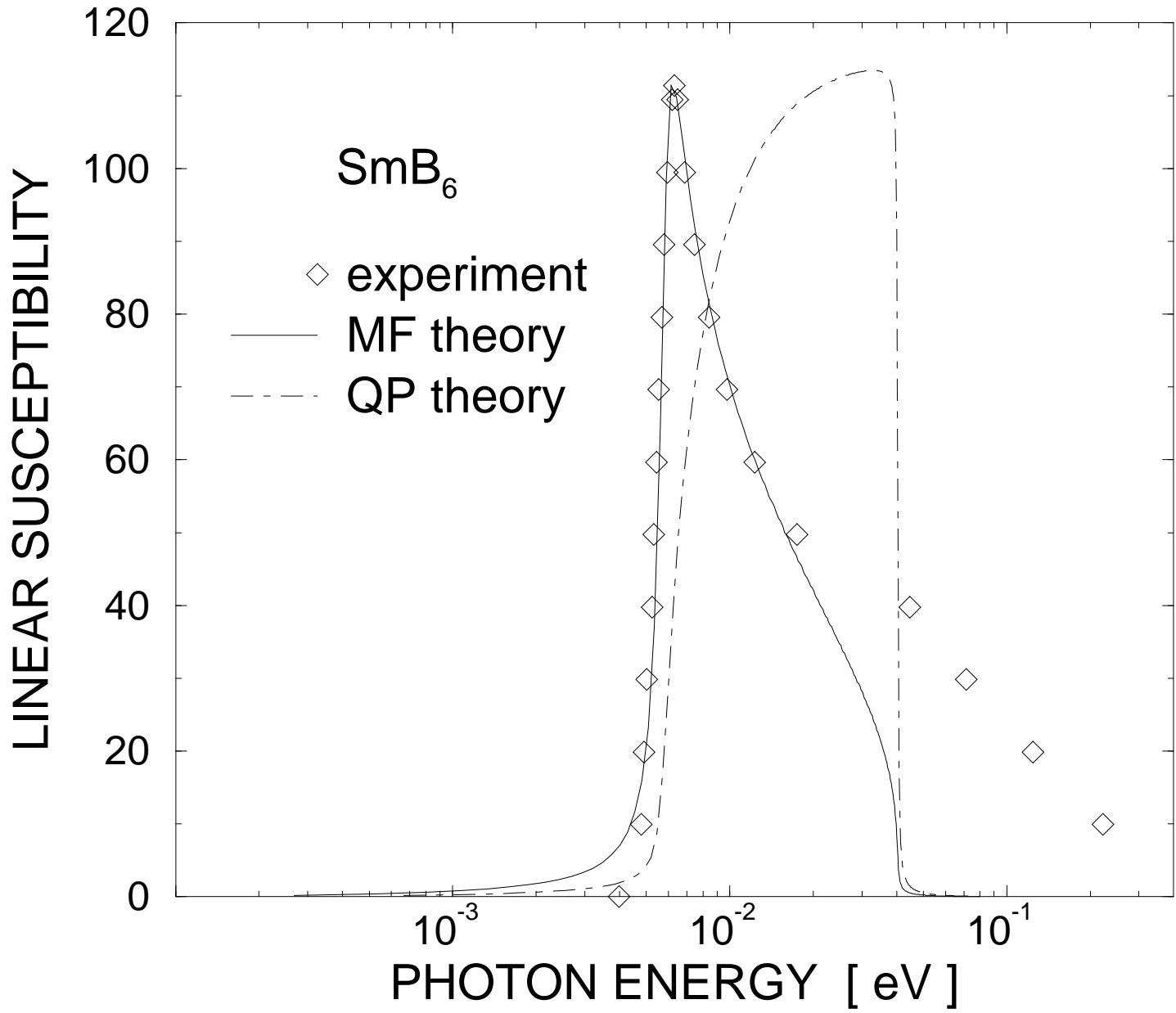
\end{document}